\documentclass[runningheads]{llncs}
\usepackage{butterma}
\idline{The final authenticated version is available online in the Proceedings of the 12th Symposium on Search-Based Software Engineering (SSBSE~2020) at Springer/SpringerLink.}

\usepackage{graphicx}
\usepackage{url}
\usepackage{xcolor}
\usepackage{boxedminipage}
\usepackage{csquotes}
\usepackage{amsmath,amssymb,dsfont,mathtools}
\usepackage{booktabs}
\usepackage{multirow}
\usepackage{float}
\usepackage{enumerate}
\usepackage{xspace}

\usepackage{hyperref}
\usepackage{cite}
\usepackage{tcolorbox}

\newcommand{\BetAndRun}{\textit{Bet and Run}\xspace}

\begin{document}
\pagestyle{headings}
\title{Bet and Run for Test Case Generation}
\titlerunning{Bet and Run for Test Case Generation}
\author{Sebastian M\"uller \and
Thomas Vogel \and
Lars Grunske}
\authorrunning{S. M\"uller et al.}
\institute{Software Engineering Group, Humboldt-Universit\"at zu Berlin, Berlin, Germany\\
\email{\{muelerse, thomas.vogel, grunske\}@informatik.hu-berlin.de}}
\maketitle
\thispagestyle{electronic}
\begin{abstract}
    Anyone working in the technology sector is probably familiar with the question: "Have you tried turning it off and on again?", as this is usually the default question asked by tech support. Similarly, it is known in search-based testing that metaheuristics might get trapped in a plateau during a search. As a human, one can look at the gradient of the fitness curve and decide to restart the search, so as to hopefully improve the results of the optimization with the next run.
    Trying to automate such a restart, it has to be programmatically decided whether the metaheuristic has encountered a plateau yet, which is an inherently difficult problem.
    To mitigate this problem in the context of theoretical search problems, the \emph{Bet and Run} strategy was developed, where multiple algorithm instances are started concurrently, and after some time all but the single most promising instance in terms of fitness values are killed. In this paper, we adopt and evaluate the \emph{Bet and Run} strategy for the problem of test case generation. Our work indicates that use of this restart strategy does \emph{not} generally lead to gains in the quality metrics, when instantiated with the best parameters found in the literature.
\keywords{Search Based Testing \and Test Case Generation \and Bet and Run.}
\end{abstract}

\section{Introduction}
	Software testing plays an important role in providing evidence for the quality of software~\cite{Orso+Rothermel2014}.
	Due to the costs and complexity of writing tests manually, automating the generation of tests	by metaheuristic search techniques is an active field of research~\cite{Orso+Rothermel2014,McMinn04,Evolutionary04}.
	A popular example for such a search-based approach to automated \textit{test case generation (TCG)} is \textit{EvoSuite}~\cite{Evosuite11,BranchDistance}. EvoSuite has shown its practical relevance by producing tests that achieve ``good levels'' of code coverage~\cite[p.\,1]{SF110.14} and detect real faults in open-source software~\cite{FraserA15}.
	The metaheuristic used by EvoSuite has evolved over time according to the state of the art, which is currently the \textit{Dynamic Many-Objective Sorting Algorithm~(DynaMOSA)}~\cite{DynaMOSA18}.
	
	Due to the stochastic nature of metaheuristics, such algorithms are approximate and therefore ``find in a reasonable computation time a solution that is as good as possible, but not necessarily optimal''~\cite[p.\,242]{Bianchi+2009}.
	Exploring the search space stochastically, a metaheuristic typically yields different results with every run, and might get trapped in an inferior part of the search space (e.g., a plateau or local optimum) which leads to suboptimal results.  As for any metaheuristic, this also applies to DynaMOSA in the context of TCG.
	
	One way to improve the search results is to restart the metaheuristic whenever it is trapped in a plateau or local optimum to try out another run~\cite{Marti2003,Lourenco2010}. However, such a restarting approach requires an appropriate strategy to be integrated in the metaheuristic that determines the trap and restarts the search~\cite{BetAndRun17}.
	To avoid such a metaheuristic-specific strategy and explicit restarts, Friedrich et al.~\cite{BetAndRun17} proposed \textit{generic \BetAndRun}, an approach that starts multiple short sample runs of a metaheuristic, evaluates the intermediate results of these runs after a certain point in time, and bets on the most promising run to continue the search until the search budget is used. This approach is generic as it is independent of the metaheuristic being used. This is in contrast to the original \BetAndRun that is intertwined with the metaheuristic~\cite{BetAndRun14}.
	Thus, by starting multiple instances of a metaheuristic and selecting the most promising one, \BetAndRun turns the problem of repeatedly restarting an instance on its head. First, there is no need to determine when a metaheuristic is trapped in an inferior part of the search space to trigger a restart. Second, the effect of repeated restarts is still obtained by starting and trying out multiple runs of the metaheuristic. The \emph{generic} \BetAndRun has been successfully evaluated on two theoretical problems (Traveling Salesperson and Minimum Vertex Cover) that are structurally different, so that Friedrich et al.~\cite{BetAndRun17} expect that \BetAndRun is generally helpful.
	
	In this paper, we adopt the idea of the \textit{generic} \BetAndRun strategy, transfer it to the TCG problem using EvoSuite/DynaMOSA, and evaluate its general feasibility for TCG and its effectiveness in comparison to the state of the art, being EvoSuite/DynaMOSA. In our \BetAndRun approach, we split the total time budget for generating test cases for a subject into two phases. In the first phase, we start multiple instances of EvoSuite/DynaMOSA concurrently, each with the same configuration but a different seed. Since \BetAndRun is generic, we can treat EvoSuite/DynaMOSA as a black box without having to change the metaheuristic. At the end of the first phase, we sample the intermediate results (test cases) in terms of the fitness score provided by EvoSuite. After the first phase, we bet on the most promising instance that continues generating test cases during the second phase, whereas all of the other instances will be terminated. Accordingly, variants of \BetAndRun in terms of how many instances will be started concurrently, and how the total time budget is split into the two phases, are possible and will be investigated in this paper. EvoSuite as the state-of-the-art approach serves as our baseline. It starts a single instance of DynaMOSA to generate test cases for a subject over the total time budget. We evaluate the effectiveness---in terms of achieved code coverage---of our \BetAndRun approach in a head-to-head comparison with this baseline, giving both approaches the same time budget to generate tests for 107 Java classes.
	
	To the best of our knowledge, no previous study has applied a generic \BetAndRun approach to a search-based software engineering problem, let alone TCG. Several TCG approaches in the literature, however, have considered restarts of search.
	The Alternative Variable Method (AVM), a variant of hill climbing, restarts the search with a randomly selected solution candidate to overcome local optima, that is, when the fitness cannot be improved~(cf.~\cite{Baars+2011,FraserAM2013,Kempka+2015}).
	Chan et al.~\cite{Chan+2006} use a restart (complete reset) as one way to forget test cases in adaptive random testing, which should reduce overheads of restricting the search space.
	Mathesen et al.~\cite{Mathesen+2019} propose a metaheuristic whose global search proposes locations where the local search is restarted if a local minimum has been found.
	Finally, a genetic algorithm has been proposed that produces the offspring (test suites) randomly in a generation rather than by evolution if the current population lacks diversity, which has an effect of restarting the search~\cite{Vogel+2019}.
	All of these approaches provide metaheuristic-specific restart strategies that are intertwined with the metaheuristic. In contrast, \BetAndRun lifts the restart strategy to a generic level that is independent of the metaheuristic being used.
		
	The main goal of this paper is to evaluate the feasibility and effectiveness of a \emph{generic} \BetAndRun approach for the TCG problem. For this purpose, we investigate the following two research questions:

	\begin{tcolorbox}[boxrule=1pt,colback=white]
	\begin{itemize}
		 \item[\textbf{RQ1}] Can \emph{generic Bet and Run} be adapted to work on the TCG problem?
		\item[\textbf{RQ2}] Does \emph{generic Bet and Run} show a significant improvement in the quality of the generated tests as measured by coverage metrics?
	\end{itemize}
	\end{tcolorbox}
	
	Accordingly, the primary contributions of this paper are:
\begin{enumerate}[(1)]
	\item As the first study, we investigate the applicability of \emph{generic Bet and Run} on the TCG problem using EvoSuite and DynaMOSA. This will answer RQ1.
	\item We conduct an empirical study to compare \BetAndRun and default EvoSuite/DynaMOSA on real-world 107 Java classes from the \emph{SF110} corpus~\cite{SF110.14}.
	\item We provide a statistical analysis of the effectiveness of \BetAndRun and EvoSuite/DynaMOSA. This will answer RQ2.
\end{enumerate}

\section{Test Case Generation with EvoSuite and DynaMOSA}
\label{sec:context}
	The context of our work is the automated \textit{test case generation (TCG)} problem that is about generating good quality tests for a given software. Particularly, we focus on generating unit tests, for which we use \textit{EvoSuite}. EvoSuite is a popular search-based ``tool that automatically generates test cases with assertions for classes written in Java code.''~\cite[p.\,416]{Evosuite11}. The metaheuristic used in EvoSuite has evolved over time from whole test suite generation~\cite{BranchDistance} to \textit{DynaMOSA}~\cite{DynaMOSA18}.

	The \textit{Dynamic Many-Objective Sorting Algorithm (DynaMOSA)}~\cite{DynaMOSA18} is the state-of-the-art many-objective genetic algorithm to solve the test case generation problem by redefining it into a many-objective problem.	Conceptually, DynaMOSA works on each statement (i.e., target to cover) of the class under test individually, instead of trying to generate a test suite for all statements simultaneously. Thus, it breaks the complex task of generating a test suite for an entire class into more manageable smaller pieces. DynaMOSA also only computes test cases for targets that can be reached immediately: All branches that are still nested below other uncovered control flow nodes are temporarily ignored. Thereby, DynaMOSA reduces the number of targets that are to be covered simultaneously and thus decreases computational complexity.
	
\section{Bet and Run for Test Case Generation}
\label{sec:bet-and-run}
	Any metaheuristic can get stuck in local optima. Once trapped, such algorithms do not tend to break free of such a plateau for a while. Friedrich et al. \cite{BetAndRun17} adopted and evaluated the approach \BetAndRun for two general theoretical computer science problems. \BetAndRun was initially presented by Fischetti and Monaci \cite{BetAndRun14} for a sequential tree search method. Friedrich et al.'s approach is no longer metaheuristic-specific and makes use of the typical multi-processing architecture of modern hardware. The strategy is characterized by first starting a number of algorithm instances that are identical in terms of call parameters but contain different starting populations. After some time only the instance that showed the most promise in terms of fitness values is then kept running.
	
	Currently in the field of TCG, no restart strategies are automatically applied. Unless the end-users intervene themselves, the optimization algorithm is simply run once, and once only. But since the typically used optimization algorithms in state-of-the-art tools such as EvoSuite are employing heuristics, the observations above hold true in this field as well. Therefore, \BetAndRun strategies are generally applicable to the \emph{Test Case Generation} problem as well.
		
	In their 2018 paper \cite{DynaMOSA18}, Panichella et al. state that DynaMOSA quickly increases test suite quality, once it starts to generate tests. Therefore, DynaMOSA is a good candidate for \BetAndRun, as we can select the most promising instance early in the search process.
		
	\begin{figure}[t]
	\begin{center}
		\includegraphics[width=.95\textwidth]{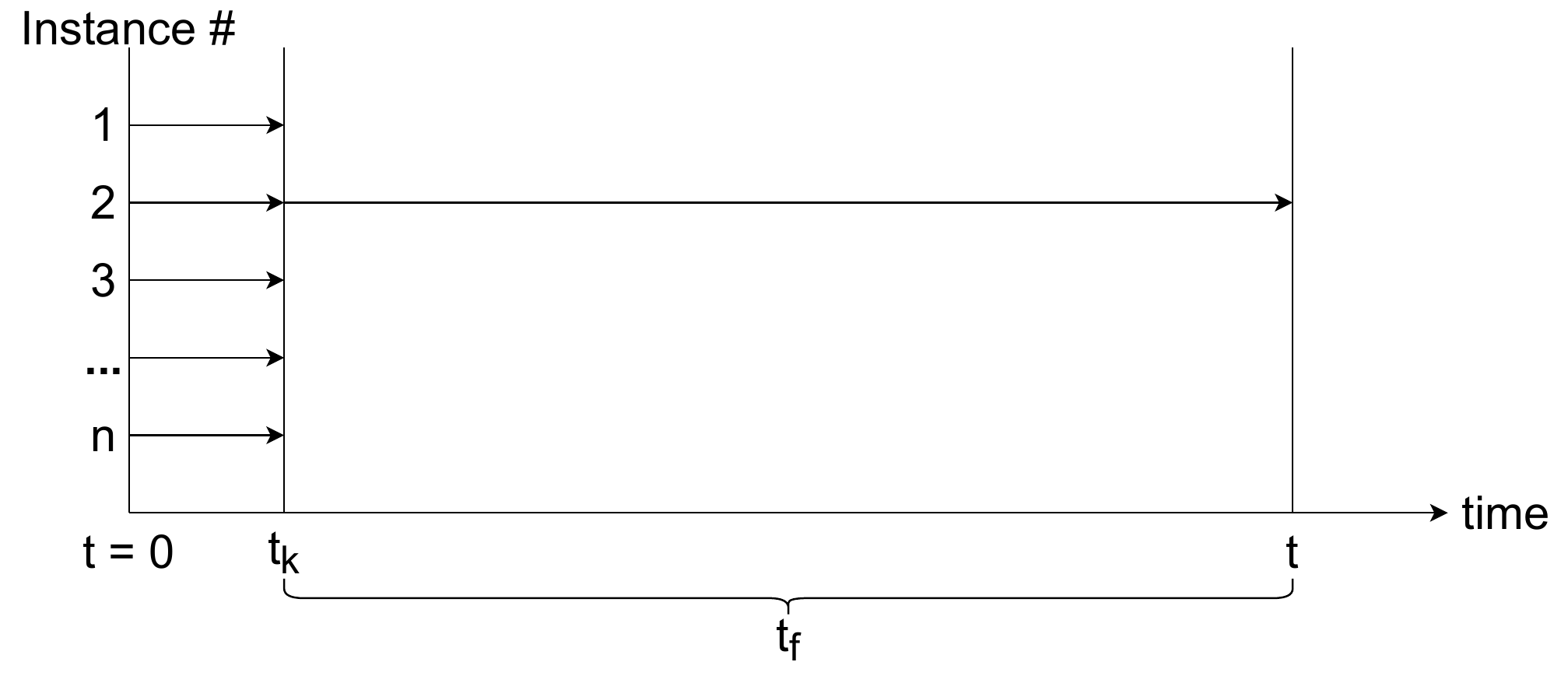}
	\end{center}
	\caption{Example: Behavior of \BetAndRun with $n$ instances and a given run time $t_{total} = t_f + n \cdot t_k$ with a maximum initial run time $t_k$ for each instance. In this case instance \#2 showed the most promise at $t_k$ and was left running.}
	\label{Fig.1}
	\end{figure}
	In our study, we adapt the general \BetAndRun approach as follows:
	\begin{definition}{\emph{Bet and Run adaptation to \emph{TCG}:}}
		\label{DefinitionBetandRun}
		Let $t_{total}$ be the full time budget available to the optimization.
		\begin{enumerate}[(i)]
			\item \emph{Starting phase.} First we start $n$ instances of DynaMOSA simultaneously, for some fixed $n$. While all typical run parameters are held fixed, the seed for the random number generator used for the optimization is varied across all $n$ instances.
			\item \emph{Evaluation phase.} After the initially selected but fixed time $t_k$ has passed, we evaluate all $n$ instances for their fitness in terms of the overall fitness score computed by EvoSuite for the tests generated so far.
			\item \emph{Elitism phase.} We then kill the least fit $n-1$ instances. If there are two or more candidates with the exact same fitness score, we continue the first of those candidates.
			\item \emph{Run phase.} Let the single remaining (and so far most promising) instance continue until $t_{total}$ is fully used up. That is, the remaining instance runs for $t_f \coloneqq t_{total} - n\cdot t_k$.
		\end{enumerate}
	\end{definition}
	This means $n$, $t_k$, and $t_{total}$ are free parameters of our adaptation of the general \BetAndRun approach. An intuitive visualization is shown in Fig. \ref{Fig.1}.
	\par Notation: Let $p$ be the percentage so that $t_k \coloneqq p \cdot t_{total}$. We then say $RESTARTS^{n}_{p}$ when we use $n$ instances of the algorithm during the \emph{starting phase} and $t_k=p\cdot t_{total}$ as the evaluation time.

\section{Implementation}
	\label{sec:development}
	In this section we provide an overview of the TCG project EvoSuite~\cite{Evosuite11}, as well as the adaptations made by us, so that we may answer RQ1. We used EvoSuite from a self-compiled \emph{1.0.7} snapshot from the \emph{git} repository\footnote{Download of EvoSuite after commit \url{https://github.com/EvoSuite/evosuite/commit/e26a6b725539370aa22976f041f3304972e201a1} on January 10, 2020.}
	This was done as the \texttt{-generateMO} API option to use the state-of-the-art DynaMOSA algorithm is not available yet in the official 1.0.6 release version of EvoSuite.
	\par To answer RQ1, the first requirement is to make use of the EvoSuite framework so as to extend its functionality with the generic \BetAndRun restart strategy. As this approach works on top of the tool itself, we do not need to change any code of EvoSuite directly. Instead, we may use it as a blackbox tool, that is, by using its command-line API we are able to get it to run tasks for us.
	\par Therefore, it suffices to concentrate on providing a runner script implementing the \BetAndRun strategy, along with a set of tools needed for saving and evaluating raw data. Apart from those scripts, all other program functionalities (i.e., genetic operators, metrics, problem encoding, etc.) are those implemented in the EvoSuite tool.
	\par In most real-world tools, there is one limitation present: One cannot simply \enquote{pause} a given algorithm run at some arbitrary point during computation, evaluate the fitness at that point, and then decide whether to continue that run or not. Instead, many tools provide a timeout function, that allows the user to specify a runtime, after which the optimization is stopped and evaluated. For implementing the \BetAndRun approach, one therefore needs to fully restart the most promising instance after the given evaluation. This, however, leads to two main strategies of coping with the tool limitation in terms of remaining runtime:
	\begin{itemize}
		\item It is possible to give the remaining instance after the full restart a maximum runtime of $t_f + 1 \cdot t_k$, in order to emulate the theoretical idea of the generic \BetAndRun approach, where \enquote{pausing} the initial $n$ instances would be possible. However, this implies that \BetAndRun has $t_k$ more computation time, when compared to the normal algorithm run without the restart strategy.
		\item A second possibility is to give the most promising instance after the full restart a maximum runtime of $t_f$. In this case, the time limit ensures, that both the normal algorithm \emph{and} \BetAndRun do not take more than $t_{total}$. 
	\end{itemize}
	In this paper, we decided to use the latter approach, so as to not give \BetAndRun any more runtime than the baseline is given. This helps to make the results between both approaches comparable.
	\par To run the prototype implementation of \BetAndRun, EvoSuite as a stand-alone, blackbox tool only requires a runner script to build the command-line API calls. This paper provides two such runner scripts: the \emph{evosuiteRunner}, and the \emph{evosuiteBARRunner}.
	Both of those runners are automatically called by the main \emph{experiments.py}.
	While the first runner simply provides an interface for running the standard EvoSuite implementation on many classes in parallel (i.e., the $RESTARTS^{1}_{100\%}$), the second one includes all changes necessary for the \BetAndRun variants. Both runners also collect data provided by EvoSuite and write the raw values to a checkpoint file per class. Additionally, both scripts are able to parse this raw data and thus compress the verbose output to the metrics.
	
	All used scripts, parameter and settings files, as well as the used EvoSuite sources are made available publicly with the online resources of this paper\footnote{\url{https://www.doi.org/10.5281/zenodo.3903206}}.
	
\section{Evaluation}
	\label{sec:evaluation}
	In this section, we present our experimental set up, lay out the methodology with which we analysed the raw data gathered, present and discuss the results of our analysis, and finally we briefly discuss threats to validity. We provide both of the interesting raw data sets with the online resources of this paper.

\subsection{General experimental set-up}
	All experiments of this paper, and the algorithms they are based upon, are implemented in either \emph{Java OpenJDK 1.8.0} or \emph{Python 3.6.10}. Additionally, for building the EvoSuite files, a \emph{Maven} installation is required. All other tool versions are listed in the online resources accompanying this paper.
	
	\paragraph{Computational resources:} The experiments were run concurrently on a \emph{Dell R920} compute server with 4 \emph{Intel Xeon E7-4880 v2} processors (60 cores) @ 2.5GHz, 1024GB RAM, and a \emph{Suse Leap 15} OS (running \emph{Linux} kernel 4.12.14). For both experimental approaches we used 40 parallel worker threads.
	
	\paragraph{Baseline Selection:}
	To investigate both the applicability (RQ1) as well as the effectiveness (RQ2) of the \BetAndRun strategy, DynaMOSA is the natural baseline to compare with. Not only is it the current state-of-the-art algorithm in the field of \emph{Test Case Generation}, it is also the algorithm extended in this study: We implemented the restart strategy using the EvoSuite tool (by use of its command-line API) and its implementation of the DynaMOSA algorithm.
	
	\paragraph{Data Set Generation:}
	Finding a set of classes that is both large enough to be representative of the state of the art in software development, as well as comparable to the studies in the field presents a certain challenge, as the adapted data sets are often not made easily available for the public.
	\par The \emph{SF100} corpus of classes is a collection of 100 Java projects from \url{www.sourceforge.net}. The data set was designed to be statistically representative. In 2014 Fraser et al. \cite{SF110.14} accounted for the fact that \emph{SourceForge} is home to a large number of older (and thus stale) projects, by revising the data set and adding the 10 most popular projects to it. This resulted in the data set now known as \emph{SF110}\footnote{\url{http://www.evosuite.org/experimental-data/sf110/}}. This benchmark has been used in many studies in the field \cite{SF110Use15,Evosuite17}, and most relevant to this study, it is also the main source of subjects for the study by Panichella et al. \cite{DynaMOSA18}, in which they introduced DynaMOSA. Thus, to be comparable to their results, we chose to adopt the \emph{SF110} benchmark. 
	As we could not find the exact list of classes used by Panichella et al. in their 2018 study, we followed their class selection algorithm:
	\begin{enumerate}
		\item Compute McCabe's cyclomatic complexity \cite{cyclomaticComplexity} for each method in the SF110 dataset using the \emph{CKJM} library\footnote{\url{http://gromit.iiar.pwr.wroc.pl/p_inf/ckjm/}}.
		\item Discard any project that only contains trivial methods (i.e., classes that only contain methods with a cyclomatic complexity of 1).
		\item Randomly sample from the remaining projects.
	\end{enumerate}
	\par Applying the above algorithm to the \emph{SF110} corpus of classes resulted in a dataset of 107 different, non-trivial classes from 68 projects as provided by the \emph{SF110} corpus of classes. The projects vary widely in size and difficulty: from a mere 18 branches on average (with 122 statements) in project \texttt{lavalamp}, to an average of 2515 branches in a medial 5887 statements in project \texttt{jmca}. This dataset was used for all experiments of this study. A complete list of all selected classes are made public with the online resources accompanying this paper.
	
\subsection{Experimental set-up for RQ1}
\label{sec:experimentsRQ1}
	\par $RESTARTS^{1}_{100\%}$ is the standard run of a given algorithm without any restarts -- and is thus ideally suited as a baseline. Friedrich et al.~\cite{BetAndRun17} have shown that $RESTARTS^{40}_{1\%}$ is the parameter set that yields the best results for both theoretical problems in their study. Of the non-trivial settings, we thus initially try $RESTARTS^{40}_{1\%}$, $RESTARTS^{20}_{2\%}$, and $RESTARTS^{8}_{5\%}$.
	
	\par Therefore, to answer RQ1 and understand whether \BetAndRun actually can work in the field of TCG, we ran an experiment using the above-mentioned computational settings, baseline, and dataset to test the four \BetAndRun strategies.
	
	\par The parameters for the \BetAndRun experiments were thus set as follows: $n = 1$, $40$, $20$, and $8$ initial runs, as well as $t_k = 100\%$, $1\%$, $2\%$, and $5\%$ of $t_{total}$. The total time budget ($t_{total}$) was set to 5 minutes. In any run of the \BetAndRun strategies $t_{total}$ is split into $n\cdot t_k$ and $t_f$ for the two phases (cf.~Def.~\ref{DefinitionBetandRun}). 
	Seeds for the initial state of the random number generator were selected at random for both the baseline and the initial phase of all employed \BetAndRun strategy executions. For the \BetAndRun experiments, we set the timeout $t_k$ for all initial instance runs via the \texttt{-Dglobal\_timeout} API parameter. On reaching time $t_k$, we evaluate each instance using the overall fitness score as provided by EvoSuite. This fitness score is an approach level (i.e., a distance) and means \enquote{lower is better}. After that evaluation, we only run the single most promising candidate by restarting that instance with the timeout now set to $t_f$, in an effort to not give \BetAndRun more runtime than the baseline. All four approaches were tested 10 times each in order to account for the pseudo-random nature of the employed DynaMOSA algorithm. A single experimental run of any given class took between approximately 60s to 5min. The entire experiment for answering RQ1 took about 3 full days per tested approach, when we used 40 concurrent worker threads.

\subsection{Answering RQ1}
	\label{ResultsRQ1}
	In our first experiment, neither \emph{$RESTARTS^{40}_{1\%}$} nor \emph{$RESTARTS^{20}_{2\%}$} managed to generate any usable data. This is true for \emph{all} 107 classes in the dataset. This problem is due to the fact that the requested test case generation time of the initial phase ($t_k$) -- as required by the respective restart strategy -- was too short for DynaMOSA to have managed to produce any tests while within the actual generation phase of EvoSuite. The restart strategy thus could not select any seed for the continued computation past $t_k$ and therefore failed. In the \emph{$RESTARTS^{8}_{5\%}$} experiment, this particular problem happened to two of the selected classes: \emph{Evaluation} of project \emph{weka}, and \emph{MethodWriter} of project \emph{jiprof}. For all other classes \emph{$RESTARTS^{8}_{5\%}$} did manage to generate at least one test suite over the ten experimental runs and thus qualified for further evaluation.
	
	\begin{tcolorbox}[boxrule=1pt,colback=white]
		\textbf{Answer to RQ1:} The \BetAndRun approach can indeed be adapted to yield usable results within the field of TCG, as shown by our experimental runs of \emph{$RESTARTS^{8}_{5\%}$}.
	\end{tcolorbox}	

\subsection{Experimental set-up for RQ2}
	Using the result from RQ1, we select \emph{$RESTARTS^{8}_{5\%}$} for the statistical comparison to the baseline. To determine the effectiveness of the selected \BetAndRun strategy, we keep the dataset, computational set-up, and the parameters of both strategies the same as they were set to in the experiment for RQ1. We do, however, in order to generate a robust statistical evaluation repeat the execution of each strategy 30 times.
	Furthermore, we also log for evaluation the following fitness metrics upon fully using the time budget for each class.
	
\paragraph{Fitness Metrics:}
	\label{par:Metrics}
	EvoSuite provides a number of different fitness metrics for each run by default:
	An overall fitness score, as well as Line, Branch, Exception, Weak Mutation, Output, Method, Method No Exception, and C Branch coverages. Additionally, we also track per class and per experimental run if EvoSuite encountered any internal errors during computation.
	\par The overall fitness score is an aggregated \emph{approach level} of the test suite to covering the remaining targets. As this approach level is a distance, it is to be read as \enquote{lower is better}. Any coverage metric is a percentage of reached coverage and all are thus to be read as \enquote{higher is better.} The according definitions used by EvoSuite may be found in \cite{Evosuite17}, as well as in the source code of the tool. The number of internal EvoSuite errors, which is also tracked, is an absolute value, thus \enquote{lower is better}.
	
	\par The full experiment for RQ2 took another 18 days. Note, that these times do not include script run times for the subsequent evaluation of the raw data gathered.

\subsection{Evaluation methodology for RQ2}
	The first step in answering RQ2, is to provide an overview over the actual behavior of the collected measures with regards to the two experiments. For \emph{mean} values given in this paper the reader can safely assume \enquote{higher is better} for every metric, with the exception of the two measures \emph{Time} and \emph{Fitness Score}. The two excluded indicators need to be handled differently:
	The \emph{Fitness Score} is an \emph{approach level} of the test suite (cf. Section~\ref{par:Metrics}), where \enquote{lower is better}. The measure \emph{Time} has only informative character and shows how long each of the final instances spent in the actual generation phase.
	\par In the subsequent statistical analysis (see Section~\ref{Sec:Analysis}), we employ the non-parametric \emph{Wilcoxon Rank Sum} hypothesis test to determine statistical significance of our results for each class and each measure taken.

\subsection{Answering RQ2}
	Generally, the results of our second experiment are structurally very similar across the different classes, as well as metrics logged. Raw data from both the \emph{baseline} and the \emph{$RESTARTS^{8}_{5\%}$} for all classes (that generated at least one test case) and for all metrics are made available in the accompanying online resources. There we provide the *.zip of all raw data gathered, as well as a *.txt file with aggregated raw data, which was used in the evaluation below in this section.
	\par We note that in the baseline experimental runs, 14 of the original 107 selected classes are rendered unusable for statistical evaluation, due to an   excessively high number of internal EvoSuite errors. Of the original set of classes in the \emph{$RESTARTS^{8}_{5\%}$} experiments, 15 are failures due to internal errors of the tool. Of all failures 12 classes fail too often for both approaches, 3 only when using DynaMOSA, and 4 only using \BetAndRun. Of those 12 classes failing for both approaches 1 class completely fails to generate \emph{any} test cases for either approach (class \emph{Evaluation} of project \emph{weka}). In total, this problem leads to 17 of the 107 classes not being eligible for statistical evaluation, apart from looking at which of the approaches was able to execute EvoSuite more stably.

\subsection{Statistical Analysis and Discussion}
\label{Sec:Analysis}
    In this section, we will present the results of the statistical analysis for the complete experiments of DynaMOSA and \emph{$RESTARTS^{8}_{5\%}$}. We obtained the $p$-values using the non-parametric \emph{Wilcoxon Rank Sum} hypothesis test, as provided by \emph{SciPy}. Note that of the 107 initially selected classes only 90 were eligible for statistical evaluation due to internal EvoSuite errors. Table~\ref{tab:dataAnalysisAggregated} shows per metrics collected the number of classes for which \BetAndRun (BAR) is (i) identical to, (ii) worse than, and (iii) better than the baseline (BL). In each column, within the brackets we also show the number of statistically significant results of the total number of classes within that field.  For instance, the first line is to be read as: For the metric \emph{Fitness Score} there were a total of 28 classes that led to identical results in both approaches. Of those 28, 28 are statistically significant. Further there were 19 classes where the baseline outperformed the \BetAndRun approach, of which 2 were statistically significant. Finally, there were 43 instances, where \BetAndRun outperformed the baseline approach, of which only 4 classes were statistically significant.
	
	\begin{table}[ht]
        \begin{center}
    	\caption{Shown here are the aggregated results of the statistical analysis over all metrics when comparing the baseline DynaMOSA (BL) and the \emph{Bet and Run $RESTARTS^{8}_{5\%}$} strategy (BAR), where BAR $=$ BL means both approaches show statistically the same result ($p$-value = 1.0), BAR $<$ BL means the \BetAndRun approach was worse than the baseline, and BAR $>$ BL means \BetAndRun was better than the Baseline. Any value inside of brackets indicate the number of statistically significant results in that field. Note that of the 107 initially selected classes only 90 were eligible for statistical evaluation due to internal EvoSuite errors.}
    	\label{tab:dataAnalysisAggregated}
    	\begin{tabular}{l@{\hskip 14pt}r@{\hskip 14pt}r@{\hskip 14pt}r@{}}
    		\toprule
    		\textbf{Metric}     			& \textbf{BAR $=$ BL}	& \textbf{BAR $<$ BL}	& \textbf{BAR $>$ BL}\\ \midrule
    		Fitness Score      			 	& 28 (28) 				& 19 (2) 				& 43 (4) \\
    		Line Coverage       			& 39 (39) 				& 13 (1) 				& 38 (3) \\
    		Branch Coverage     			& 35 (35) 				& 16 (2) 				& 39 (4) \\
    		CBranch Coverage    			& 36 (36) 				& 16 (2) 				& 38 (4) \\
    		Method Coverage     			& 78 (78) 				&  6 (0) 				&  6 (0) \\
    		Method no Exception Coverage   	& 69 (69) 				& 11 (0) 				& 10 (1) \\
    		Exception Coverage      		& 80 (80) 				&  5 (0)	 			&  5 (0) \\
    		Output Coverage     			& 57 (57) 				& 12 (0) 				& 21 (1) \\
    		Weak Mutation Coverage      	& 38 (38) 				& 19 (0) 				& 33 (1) \\ \bottomrule
    	\end{tabular}
    	
    	\end{center}
    \end{table}
    \par Table \ref{tab:dataAnalysisAggregated} shows that neither approach was significantly better than the other overall. In fact, in most cases the $p$-value either does not reach statistical significance or even indicates that both approaches reach statistically the exact same result. Over all 9 metrics and all 90 evaluated classes there were only 25 results that show any statistically significant \emph{difference} between the two approaches. Of those, \BetAndRun was 18 times significantly better than the baseline, and 7 times significantly worse (cf. Table~\ref{tab:dataSignifClasses}). In contrast, over all 9 metrics and all 90 classes, we see 460 instances in which both approaches yield the exact same result. This means, when looking at \emph{all} 810 results (i.e., 9 metrics x 90 classes) 785 instances are either only marginally different, or yield the exact same result.

 	\begin{table}[ht]
 		\caption{Shown here are the classes that reached a significant difference when comparing the baseline DynaMOSA and the \emph{Bet and Run $RESTARTS^{8}_{5\%}$} strategy.}
 		\begin{tabular}{@{}ccc@{}}
 			\toprule
 			\textbf{Metric}     & \textbf{\begin{tabular}[c]{@{}l@{}}Stat. sig.\\ worse classes \end{tabular}} & \textbf{\begin{tabular}[c]{@{}l@{}}Stat. sig.\\ better classes \end{tabular}} \\ \midrule
 			Fitness Score       & \begin{tabular}[c]{@{}l@{}}99\_newzgrabber: Downloader\\ 12\_dsachat: Handler \end{tabular} & \begin{tabular}[c]{@{}l@{}}19\_jmca: JMCAAnalyzer\\ 61\_noen: ProbeInformation\\82\_ipcalculator: BinaryCalculate\\ 86\_at-robots2-j: RobotRenderer \end{tabular}\\
 			Line Coverage       & 12\_dsachat: Handler & \begin{tabular}[c]{@{}l@{}}19\_jmca: JMCAAnalyzer\\ 82\_ipcalculator: BinaryCalculate\\ 86\_at-robots2-j: RobotRenderer \end{tabular}\\
 			Branch Coverage     & \begin{tabular}[c]{@{}l@{}}39\_diffi: IndexedString\\ 12\_dsachat: Handler \end{tabular} & \begin{tabular}[c]{@{}l@{}}19\_jmca: JMCAAnalyzer\\ 61\_noen: ProbeInformation\\ 82\_ipcalculator: BinaryCalculate\\ 86\_at-robots2-j: RobotRenderer \end{tabular} \\
 			CBranch Coverage    &\begin{tabular}[c]{@{}l@{}}39\_diffi: IndexedString\\ 12\_dsachat: Handler \end{tabular} & \begin{tabular}[c]{@{}l@{}}19\_jmca: JMCAAnalyzer\\ 61\_noen: ProbeInformation\\ 82\_ipcalculator: BinaryCalculate\\ 86\_at-robots2-j: RobotRenderer \end{tabular}\\
 			Method Coverage     & - & - \\
 			Met. no Exc. Cov.   & - & 86\_at-robots2-j: RobotRenderer\\
 			Exception Cov.      & - & - \\
 			Output Coverage     & - & 50\_biff: Scanner\\
 			Weak Mut. Cov.      & - & 61\_noen: ProbeInformation\\ \bottomrule
 		\end{tabular}
 		\label{tab:dataSignifClasses}
 	\end{table}
    \par We argue that this is indicative of a limitation of the approach: In a real-world setting in the TCG context, runs typically are strictly limited in their computation times. This strict and more importantly \emph{short} time budget for the genetic algorithm appears to mitigate the plateauing problem quite effectively. Furthermore, the use of the state-of-the-art DynaMOSA algorithm with its optimization mechanisms that work on minimizing the initial number of coverage targets (i.e., only search solutions for branches that can actually be covered at this time) help to further decrease the potential of getting stuck in a local optimum.
    \par  Panichella et al. state in their 2018 paper \cite{DynaMOSA18} that DynaMOSA quickly increases test suite quality, once it starts to generate tests. We can confirm this behavior: In our early experiments for RQ1, we saw that after a short initialization phase the algorithm managed to quickly improve fitness scores across all metrics for the population.
    We believe this quick improvement phase at the beginning is due to the fact that DynaMOSA only works on targets that can possibly be covered currently, instead of working immediately on every single target (i.e., the mechanism that the algorithm does not start working on targets that are still restricted by higher level, yet uncovered targets). This behavior is also a possible explanation for what we stated above: Neither \emph{$RESTARTS^{40}_{1\%}$}, nor \emph{$RESTARTS^{20}_{2\%}$} managed to generate any tests. When we are using these settings, no candidate instance can be chosen to survive past the decision time $t_k$, as $t_k$ was too short to get past the above-mentioned initialization phase. We note that while \emph{$RESTARTS^{40}_{1\%}$} was shown in Friedrich et al. \cite{BetAndRun17} to be the best strategy, we cannot confirm this finding here. Moreover, we also cannot safely confirm quality increases in the test suite generated by EvoSuite. In fact, we would argue that the additional computational overhead is indeed an argument against usage of that particular restart strategy in the field of \emph{Test Case Generation}, when also looking at the achieved results.
	\par We have, however, seen a significant increase ($p = 0.002930$) in tool stability over all selected classes, when employing the \BetAndRun strategy \emph{$RESTARTS^{8}_{5\%}$}. While the \emph{baseline} showed a total of 578 internal errors of the tool, the \BetAndRun approach showed 500 errors across all classes. This difference of exactly 78 errors occurred over all 3210 experimental runs per approach (30 repetitions x 107 classes). That means in 15.6\% of all experimental runs, the EvoSuite tool encountered errors during the \emph{$RESTARTS^{8}_{5\%}$} experiments, in comparison with 18.0\% in the normal behavior EvoSuite experiments. This difference can be explained by the transience of internal EvoSuite errors, where a simple restart may fix the problem thanks to the selection of a new random seed value. As these transient internal tool errors usually happen very early during computation and since the \BetAndRun restart strategy never selects any seed value that showed an error, this is expected behavior of the employed strategy. We assume the rather high number of internal errors by EvoSuite is due to the fact that we employed a self-compiled (and thus potentially unstable) tool snapshot, instead of an actual release of the software. We therefore argue, that unless tool stability is of particular importance to the use case, the statistically significant gains in this metric \emph{do not} outweigh computational overhead needed for those increases.
	
	\begin{tcolorbox}[boxrule=1pt,colback=white]
		\textbf{Answer to RQ2:} Apart from statistically significant increases in tool stability, we cannot confirm any gains in the remaining nine metrics when using \BetAndRun with the best parameters found in literature. In fact, we question the effectiveness of \BetAndRun in the field of TCG, as time constraints and highly specialized algorithms (e.g., DynaMOSA) in this field effectively mitigate the typical plateauing problem of metaheuristics.
	\end{tcolorbox}
		
\subsection{Threats to Validity and Future Work}
	In this paper, we rely on a search-based approach with a restart strategy to solve the Test Case Generation Problem. As such we see the following potential threats to the validity of our work.

\paragraph{Internal:}
	In all genetic algorithms, a fair amount of (pseudo-)randomness is involved in the generation of their respective results. In DynaMOSA, the randomness problem is somewhat alleviated by the use of the archiving function -- a test case covering some new target is never lost due to future mutations again.
	Additionally, we try to mitigate this threat by (i) sampling 30 runs per selected class and evaluating all results only in the average case; as well as (ii) providing online the seed values for all runs, both in the normal EvoSuite and the \BetAndRun experiments (in the latter we provide the seeds that showed the most promise at time $t_k$). The seed values are made available in the aggregated result data files of the online resources. Furthermore, there exists the possibility of faults within our strategy implementation. We tried to minimize this threat by use of standard components (such as the EvoSuite implementation of DynaMOSA or the \emph{SciPy} libraries for the statistical testing) wherever possible. All implementations, codes, and scripts are made available for inspection, review, and validation with the online resources accompanying this study.
	
\paragraph{External:}
	Our evaluation is based upon the benchmark data set \enquote{\emph{SF110 Corpus of Classes}} as provided by Fraser et al. \cite{SF110.14}. While Fraser et al. argue that they selected 100 statistically representative \emph{SourceForge} projects and even further enhanced this set with another 10 \enquote{popular} projects, the possibility that the heterogeneous nature of software is indeed not fully accounted for still exists: The restart strategy might behave differently for a different set of \emph{real} classes. We leave for future work to revisit the use of the restart strategy with a different data set.
	
\paragraph{Construct:}
	The first threat in this category that we can see is the possibility of the experiments not yet having encountered any plateaus. This threat could only be mitigated by redoing the entire set of experiments with a much higher timeout. However, this would go against real world usage of the EvoSuite tool. In practical settings, the tool is used to \emph{quickly} generate a test suite for project classes -- extremely high timeouts are thus the exact opposite of the intended use. We will, however, leave for future work a confirmation study with those increased timeouts: It is a matter of running the provided implementation with a new set of parameters. A second threat of this family is that while we selected restart strategy parameters that were shown to be \enquote{good} in the study by Friedrich et al. \cite{BetAndRun17}, \enquote{good} in the context of TCG might be something else entirely. In our case, it may be the case that most of $t_k$ is in fact used for tool initialization, and no significant test generation is performed so as to correctly estimate the "best" performing one. Similarly to the first threat to construct validity, a parameter tuning study with a new set of parameters is left for future work, as even more parameters for the restart strategy would go beyond the scope of this initial study. And thirdly, by forcing us to restart the most promising instance completely, the implementation limited us to choose between either giving the remaining instance in the run phase a maximum run-time of (i) $t_f + t_k$, or (ii) $t_f$. We chose the latter, so that both \BetAndRun and the baseline do not use more than $t_{total}$. This leads to the evaluated instance having $p\%$ less run-time, while the total run-time of the \BetAndRun approach is $100\%$ of $t_{total}$. Had we used approach (i), however, \BetAndRun would have had a total run-time of $t_{total}+t_k$.

\section{Conclusion}
	\label{sec:conclusion}
	In this study, we provided a tool to run and evaluate the generic restart strategy \BetAndRun in the context of Test Case Generation using EvoSuite and the state-of-the-art DynaMOSA algorithm. To the best of our knowledge this was the first study that applies the generic \BetAndRun approach to this field. Our work indicates that use of a restart strategy instantiated with the best parameters found in the literature does \emph{not} generally lead to gains in the quality metrics: Not a single metric was improved by the \BetAndRun restart strategy when compared with the EvoSuite/DynaMOSA as the baseline. In fact, for most selected classes, both approaches showed statistically the exact same results.
	\par Only the stability of runs showed statistically significant improvements when \BetAndRun was employed, as the number of internal EvoSuite errors decreased in the final generation process. However, from what we saw in our experiments, most of the internal EvoSuite errors are transient: In a real-world setting where the tool did not generate a test suite immediately, it is more than likely that a second (manually started) run with the same parameters would fix the problem.
	\par Our results indicate, that (contrary to the promising results in the 2017 study by Friedrich et al.~\cite{BetAndRun17}) the restart strategy \BetAndRun is \emph{not} suited for improving the quality of automatically generated test suites using EvoSuite and its state-of-the-art DynaMOSA implementation considering the best \BetAndRun parameters found in literature.
%
% ---- Bibliography ----
\bibliographystyle{splncs04}
\bibliography{bibliography}

\end{document}